\documentclass[12pt,showpacs,showkeys,amsmath,amssymb]{revtex4}
\usepackage{amsmath,amsfonts,amsthm,amscd,amssymb,latexsym}
\usepackage{bm}
\usepackage{dcolumn}
\usepackage{graphicx}
\usepackage{epstopdf}
\usepackage{color}
\usepackage{epsf}
\usepackage{epsfig}
\usepackage{graphicx, epic, eepic, color}
\usepackage{subfig}
\usepackage{braket}
\begin{document}

\title{McVittie-Plummer Spacetime: Plummer Sphere Immersed in the FLRW Universe}

\author{Javad \surname{Tabatabaei}$^{1}$}
\email{smj_tabatabaei@physics.sharif.edu}
\author{Shant \surname{Baghram}$^{1,2}$}
\email{baghram@sharif.edu}
\author{Bahram \surname{Mashhoon}$^{1,3,4}$} 
\email{mashhoonb@missouri.edu}

\affiliation{
$^1$Department of Physics, Sharif University of Technology, Tehran 11155-9161, Iran\\
$^2$Research Center for High Energy Physics, Department of Physics, Sharif University of Technology, Tehran 11155-9161, Iran\\
$^3$School of Astronomy,
Institute for Research in Fundamental Sciences (IPM),
Tehran 19395-5531, Iran\\
$^4$Department of Physics and Astronomy,
University of Missouri, Columbia,
Missouri 65211, USA
}

\date{\today}

\begin{abstract}
The McVittie-Plummer spacetime is a spherically symmetric inhomogeneous cosmological model that represents a spherical star system embedded in a standard FLRW cosmological model. We study the main physical properties of this gravitational field. Regarding the interplay between the physics of the local system and the expanding background, we employ the Misner-Sharp mass-energy function to show that there is a relatively weak time-dependent general relativistic coupling between the astrophysical system and the background FLRW cosmological model.  The coupling term is proportional to the inverse of the scale factor and decreases as the universe expands. 
\end{abstract}


\maketitle

\section{Introduction}

The universality of the gravitational interaction implies that every massive particle is subject to the gravitational influence of the whole mass-energy content of the universe. Therefore, the relative motion of free neighboring test particles should, in general, be affected by the cosmic tidal field.  Approximating the distribution of matter in the universe via the standard Friedmann-Lema\^itre-Robertson-Walker (FLRW) cosmological models or the inhomogeneous Lema\^itre-Tolman-Bondi (LTB) models, a detailed examination has revealed that the influence of the recession of galaxies on the local solar system dynamics is negligibly small; see, for instance,~\cite{Cooperstock:1998ny, Mashhoon:2007qm, Faraoni:2007es, Kopeikin:2012by, Kopeikin:2013am, Iorio:2012wva, Spengler:2021vxy} and the references cited therein.  The relative tidal acceleration is in the lowest order proportional to the spacetime curvature multiplied by the relative distance; nevertheless, the influence of cosmic expansion on the scale of a galaxy appears to be insignificant as well. For interesting considerations regarding the interplay between the dynamics of clusters of galaxies and the expansion of the universe, see~\cite{Nandra:2011ug, Nandra:2011ui, Nandra:2013jga, Benisty:2024tlv} and the references cited therein. 

It is important to study the connection between local dynamics and the expansion of the universe within the context of exact solutions of the general theory of relativity.  To this end, we consider the gravitational field of a Plummer sphere~\cite{Plummer} embedded in the FLRW background as a simple generalization of McVittie spacetime~\cite{McVittie:1933zz}. Here, the Plummer sphere represents a gravitationally bound distribution of matter which could be a dwarf spherical galaxy or a spherical dark matter halo possibly containing a galaxy. We investigate the main physical aspects of the resulting McVittie-Plummer spacetime as an inhomogeneous cosmological model~\cite{Krasinski:1997yxj, Plebanski:2006sd, Bolejko:2009pvd}. Within the simplified framework of the McVittie-Plummer model, we show via the Misner-Sharp mass-energy function~\cite{Misner:1964je, HeMi, CaMc} that there is a general relativistic coupling between the Plummer sphere and the expanding background and the strength of the coupling decreases as the universe expands. 

Before proceeding to a detailed description of the McVittie-Plummer model, it is necessary to point out that a substantial body of literature exists on the influence of cosmic expansion on local physics and the present work only addresses one aspect of the general problem; in this connection, see~\cite{Bonnor, Kaloper:2010ec, Lake:2011ni, Nolan:2014maa, Nolan:2017rtj, Perlick:2018iye, Faraoni:2018xwo, Gaur:2023hmk} and the references therein for a more complete list of references. 

In 1911, Plummer~\cite{Plummer} introduced a simple model for the distribution of matter in globular star clusters. The Newtonian gravitational potential for the Plummer sphere is given by
\begin{equation}\label{I1}
\Phi_{\rm P} (r) = - \frac{GM}{(r^2+b^2)^{1/2}}\,,
\end{equation}
where $b$ is the Plummer radius and $M$ is the total mass of the cluster such that $GM\ll c^2 b$. More generally, for $b/r \to 0$, we recover the gravitational potential for a point mass $M$, while for $r \gg b > 0$, we have the asymptotic Newtonian potential for a system of total mass $M$. From the Poisson equation of Newtonian gravitation,  
\begin{equation}\label{I2}
\nabla^2 \Phi_{\rm P} = 4 \pi G \rho_{\rm P}\,, 
\end{equation}
we find the density of matter in the cluster, namely, 
\begin{equation}\label{I3}
\rho_{\rm P}(r) = \frac{3\,b^2 M}{4 \pi} \frac{1}{(r^2+b^2)^{5/2}}\,.
\end{equation}
The mass within a sphere of  radius $r$ is given in this case by
\begin{equation}\label{I4}
M(< r) = \frac{M r^3}{(r^2+b^2)^{3/2}}\,,
\end{equation}
where $M(<r) \to M$ for $r \to \infty$. 

For $r \gg b$, the mass density $\rho_{\rm P}(r)$ falls off as $r^{-5}$, which is too steep for current models of globular clusters; nevertheless, the simple Plummer model is still quite useful in treating the stellar density in dwarf galaxies and in dealing with point particles in $N$-body simulations.  

To render the present paper essentially self-contained, we include brief descriptions of the relativistic Plummer model in Section II and the standard spatially homogeneous and isotropic cosmological models in Section III.  In Section IV, we present the McVittie-Plummer spacetime, which is a generalization of McVittie spacetime and represents a simple embedding of the Plummer sphere in the FLRW cosmological background.  We explore the main physical aspects of this gravitational field in Sections V and VI. Section VII contains a brief discussion of our results. 

% @@@@@@@@@@@@@@@@@@@@@
% @@@@@@@@@@@@@@@@@@@@@
% @@@@@@@@@@@@@@@@@@@@@

\section{Relativistic Plummer Model}

In Newtonian physics,  space is  Euclidean and isotropic. Hence, to generalize Plummer's Newtonian model to the relativistic regime, we adopt isotropic coordinates and write the spacetime metric as
\begin{equation}\label{R1}
ds^2 = -  \left(\frac{1-\phi}{1+\phi}\right)^2 dt^2 + (1+\phi)^4 \delta_{ij}\,dx^i \, dx^j\,,
\end{equation}
where $\phi := - \tfrac{1}{2}\,\Phi_{\rm P}$, namely, 
\begin{equation}\label{R2}
\phi = \frac{1}{2}\,\frac{M}{(r^2+b^2)^{1/2}}\,.
\end{equation}
Here, the Plummer radius $b > 0$ is a constant  and $r = (\delta_{ij}\,x^i \, x^j)^{1/2}$ is the radial coordinate.  Moreover, $x^\mu = (ct, \mathbf{x})$, Greek indices run from 0 to 3, while Latin indices run from 1 to 3. The signature of the metric is +2 and we use units such that $c = G = 1$, unless specified otherwise.  Various aspects of the relativistic Plummer model have been the subject of previous investigations, see~\cite{BuD, Facke, GlaMa}. 

Einstein's field equations are~\cite{Einstein}
\begin{equation}\label{R3}
G_{\mu \nu} + \Lambda\, g_{\mu \nu}=\kappa\,T_{\mu \nu}\,, 
 \end{equation}
 where $G_{\mu \nu}$ is the Einstein tensor  
\begin{equation}\label{R4}
 G_{\mu \nu} := R_{\mu \nu}-\frac{1}{2} g_{\mu \nu}\,R\,.
 \end{equation} 
We denote the symmetric energy-momentum tensor of matter by $T_{\mu \nu}$; moreover,  $\Lambda$ is the cosmological constant and $\kappa:=8 \pi G/c^4$.  
Let us return to metric~\eqref{R1} and note that it satisfies the gravitational field equations of general relativity with $\Lambda = 0$ and a $T_{\mu \nu}$ given by a perfect fluid with energy density $\mu_{\rm P}$ and pressure $p_{\rm P}$ such that 
\begin{equation}\label{R5}
T_{\mu \nu}=\mu_{\rm P}\,u_\mu\,u_\nu+p_{\rm P}\,(g_{\mu \nu}+u_\mu\,u_\nu)\,,
\end{equation}
where $u^\mu$ is the 4-velocity vector of the perfect fluid.  We assume the fluid particles are  spatially at rest, namely, 
\begin{equation}\label{R6}
 u^\mu = dx^\mu/d\tau_{\rm P} = \frac{1+\phi}{1-\phi}\,\delta^\mu _0\,, 
\end{equation} 
where $\tau_{\rm P}$ is the proper time. The motion of the fluid is shear free in all of the solutions discussed in  this paper.

It follows from the field equations that the energy density is given by
\begin{equation}\label{R7}
\mu_{\rm P} = \frac{24\,b^2}{\pi\,M^4}\frac{\phi^5}{(1+\phi)^5}\,,
\end{equation}
which for $\phi \ll 1$ reduces to $\rho_{\rm P}$, as expected. Moreover, the pressure is given by
\begin{equation}\label{R8}
p_{\rm P} = \frac{1}{3}\frac{\phi}{1-\phi}\,\mu_{\rm P}\,.
\end{equation}
This formula corrects the result given in~\cite{GlaMa} by restoring the energy density $\mu_{\rm P}$, which is absent in~\cite{GlaMa}  due to a typographical error.  Let us note that $0 < p_{\rm P} < \tfrac{1}{3}\,\mu_{\rm P}$ for  $0 < \phi < \tfrac{1}{2}$.

It turns out that this solution of general relativity is the relativistic generalization of  the Emden polytrope of index 5. The polytrope equation of state is $p = \kappa_0\, \mu^{1+1/n}$, where $\kappa_0$ is a constant and $n$ is the polytrope index.
The equation of state for this solution has been discussed by Buchdahl~\cite{BuD}. 

In spherically symmetric spacetimes, it is possible to define invariantly a certain function $m(t, r)$ that defines the amount of mass-energy within a radius $r$ at time $t$ and has interesting properties~\cite{Misner:1964je, HeMi, CaMc}. We, therefore, turn to a description of this Misner-Sharp mass~\cite{Misner:1964je} for the relativistic Plummer sphere. 

\subsection{Misner-Sharp Mass}

The spacetime metric for a spherically symmetric spacetime in general depends on two functions of time and radial coordinate, while the radial motion of the source of the gravitational field remains arbitrary. On the other hand, if we demand that the material source of the field be stationary, namely, remain at rest in space, then we need three functions to describe the spacetime metric. Using spherical  polar coordinates, we can write the spacetime metric in comoving coordinates $x^\mu = (t, r, \vartheta, \varphi)$  as
\begin{equation}\label{R9}
ds^2 = - A^2(t, r)\, dt^2 + B^2(t, r)\, dr^2 + R^2(t, r)\, (d\vartheta^2 + \sin^2\vartheta\, d\varphi^2)\,.
\end{equation}
The magnitude of the azimuthal Killing vector field is an invariant; hence, it is possible to define invariantly the areal radius $R(t, r)$. The Misner-Sharp mass-energy function $m$~\cite{Misner:1964je, HeMi, CaMc} is then given by
\begin{equation}\label{R10}
\frac{2m}{R^3}  = \mathcal{K}\,, \qquad  \mathcal{K} = \frac{R_{\mu \nu \rho \sigma}\,\vartheta^\mu\, \varphi^\nu\, \vartheta^\rho\, \varphi^\sigma}{\vartheta^\alpha\,\vartheta_\alpha\,\varphi^\beta\,\varphi_\beta}\,.
\end{equation}
Here, $\mathcal{K}$ is the surface sectional curvature, while $\vartheta^\mu = \delta^\mu_2$ and $\varphi^\mu = \delta^\mu_3$ are two linearly independent tangent vectors that are orthogonal to each other and to the radial direction. In terms of metric~\eqref{R9}, we find the Misner-Sharp mass $m(t, r)$ is given explicitly by~\cite{GlaMa, Mashhoon:1979tt}
\begin{equation}\label{R11}
m(t, r) = \frac{1}{2} R \left[ 1 + \left(\frac{1}{A}\frac{\partial R}{\partial t}\right)^2 -  \left(\frac{1}{B}\frac{\partial R}{\partial r}\right)^2\right]\,.
\end{equation}
We can now use this general formula to compute the Misner-Sharp mass-energy function for various comoving configurations. 

The Misner-Sharp (MS) mass for the relativistic Plummer model is given by $m_{\rm P}$, 
\begin{equation}\label{R12}
m_{\rm P} = M(< r) + \frac{1}{2}\,\frac{GM}{c^2}\,\frac{Mb^2\,r^3}{(r^2+b^2)^3}\,,
\end{equation}
which consists of a purely Newtonian part $M(< r)$ given by Eq.~\eqref{I4} plus a small general relativistic correction. For $b \to 0$, $m_{\rm P} \to M$, as expected for the Schwarzschild black hole. It will turn out that when the Plummer sphere is embedded in the FLRW universe, the general relativistic correction in Eq.~\eqref{R12} becomes time dependent and decreases with the expansion of the universe as the inverse of the FLRW scale factor.

% @@@@@@@@@@@@@@@@@@@@@
% @@@@@@@@@@@@@@@@@@@@@
% @@@@@@@@@@@@@@@@@@@@@

\section{FLRW Models}

For the sake of completeness, we briefly describe in this section  the standard FLRW cosmological models given by the metric
\begin{equation}\label{H1}
 ds^2 = -dt^2 + \frac{a^2(t)}{K^2(r)}\,\delta_{ij}\,dx^i\,dx^j\,, 
\end{equation} 
where $a(t)$ is the scale factor, 
\begin{equation}\label{H2}
K(r)  = 1 + \frac{k}{4R_0^2 }\, r^2\,, \qquad  r^2 = \delta_{ij}\, x^i x^j\,, 
\end{equation}  
and $k = 1, -1$, or $0$,  for the closed, open, or flat model, respectively. Here, $R_0$ is a characteristic cosmological length scale. 

The preferred observers in the spatially homogeneous and isotropic FLRW background are fixed in space and have adapted orthonormal tetrad frame field $\lambda^{\mu}{}_{\hat \alpha}$ such that 
\begin{equation}\label{H3}
\lambda^{\mu}{}_{\hat 0} = \delta^\mu _0\,, \qquad \lambda^{\mu}{}_{\hat i} = \frac{K(r)}{a(t)} \,\delta^\mu_i\,.
\end{equation}  
This tetrad frame field is parallel transported along the geodesic world lines of the preferred observers, each with 4-velocity vector $\mathfrak{u}^\mu = \lambda^{\mu}{}_{\hat 0} =  \delta^\mu_0$ and proper time equal to cosmic time $t$. 

The source of the gravitational field is a perfect fluid of uniform density and pressure comoving with the preferred observers. The field equations in this case can be worked out using metric~\eqref{H1} and the nonzero components of the  Einstein tensor in the absence of the cosmological constant are given by 
\begin{equation}\label{H4}
G_{00} = \kappa \,\mu_{\rm FLRW} = 3\, \left(\frac{\dot{a}}{a}\right)^2 + 3\,\frac{k}{a^2\,R_0^2}\,
\end{equation}
and
\begin{equation}\label{H5}
G_{ij} = \kappa\, p_{\rm FLRW}\, \frac{a^2(t)}{K^2(r)}\,\delta_{ij} = -\frac{1}{K^2}(2a\,\ddot{a} + \dot{a}^2 + k/R_0^2)\,\delta_{ij}\,.
\end{equation}

Let us define the Hubble and deceleration parameters $H$ and $q$,
\begin{equation}\label{H6}
H = \frac{\dot{a}}{a}\,, \qquad  q H^2 = - \frac{\ddot{a}}{a}
\end{equation}
where $\dot{a} = da/dt$, etc.  We note that 
\begin{equation}\label{H7}
\dot{H} := \frac{dH}{dt} = \frac{\ddot{a}}{a} - H^2 = -H^2 (1+q)\,,
\end{equation}
where $q = -1$ for de Sitter's cosmological model.  It is estimated that $q_0 \approx -0.55$ at the present epoch ($t = t_0$)~\cite{Dodelson:2003ft,Weinberg:2008zzc,Amendola:2015ksp,Baumann:2022mni}. 

It is useful to define $\Pi_1$ and $\Pi_2$, 
 \begin{equation}\label{H8}
\Pi_1 := qH^2= 4 \pi (p_{\rm FLRW} + \tfrac{1}{3}\,\mu_{\rm FLRW})\,, \qquad \Pi_2 = H^2 + \frac{k}{a^2\,R_0^2} = \frac{8\pi}{3}\,\mu_{\rm FLRW}\,,
\end{equation}
which we employ in Eq.~\eqref{F2} of Section VI. 

% @@@@@@@@@@@@@@@@@@@@@
% @@@@@@@@@@@@@@@@@@@@@

\section{McVittie-Plummer Spacetime}

The metric 
\begin{equation}\label{M1}
 ds^2 = - \left[\frac{1-\phi K^{1/2}(r)/a(t)}{1+\phi K^{1/2}(r)/a(t)}\right]^2 \,dt^2 + \frac{a^2(t)}{K^2(r)}\,[1+\phi K^{1/2}(r)/a(t)]^4\,\delta_{ij}\,dx^i\,dx^j\, 
\end{equation} 
is a nonlinear superposition of the relativistic Plummer model and the FLRW model such that it satisfies the field equations of general relativity for an appropriate energy density $\mu$ and pressure $p$ of a perfect fluid source. Here, $\phi$ and $K$ are given by Eqs.~\eqref{R2} and~\eqref{H2}, respectively, and $r = (\delta_{ij} x^i x^j)^{1/2}$ is the radial coordinate.  The solution~\eqref{M1} as well as its charged generalization was first found implicitly in~\cite{GlaMa, Mashhoon:1979tt}. That is, Eq.~\eqref{M1} is a member of the general class of solutions published in~\cite{GlaMa, Mashhoon:1979tt}, which were devoted to the problems of gravitational collapse and the resulting black holes.  In contrast, the present work is  dedicated instead to the cosmological aspects of a dwarf spherical galaxy embedded in an expanding universe. 

For $b\to 0$, $\phi \to M/(2r)$ and metric~\eqref{M1} reduces to the McVittie metric~\cite{McVittie:1933zz}. Just as in the case of the McVittie model,  the source of the gravitational field can be assumed to be a perfect fluid
\begin{equation}\label{M2}
T_{\mu \nu}=\mu\,U_\mu\,U_\nu+p\,(g_{\mu \nu}+U_\mu\,U_\nu)\,,\qquad U^\mu = (-g_{tt})^{-1/2}\, \delta^\mu_0\,,
\end{equation}
where $U^\mu$ is the timelike unit 4-velocity of the perfect fluid such that the fluid particles are spatially at rest in metric~\eqref{M1}, in conformity with the comoving assumption in the standard cosmological models. The Einstein field equations in this case are  given by  $G_{\mu}{}^{\nu} = 8\pi\,T_\mu{}^\nu$, where $T_\mu{}^\nu = \text{diag}(-\mu, p, p, p)$ and the cosmological constant is assumed to be zero ($\Lambda = 0$). The gravitational field equations reduce to the two formulas for $\mu$ and $p$. To this end, let us define 
\begin{equation}\label{M3}
\chi := \frac{k}{4 R_0^2}\,, \qquad k = -1, 0, 1\,, 
\end{equation}
\begin{equation}\label{M4}
\mathbb{X}(r) := \frac{M}{2 \phi K^{1/2}} = \left(\frac{r^2 + b^2}{\chi\, r^2 +1}\right)^{1/2}\,,
\end{equation}
where only positive square roots are considered throughout. Then, we find 
\begin{equation}\label{M5}
\mu = \frac{3 H^2}{8 \pi} +  \frac{3}{4 \pi a^2} \mathbb{A}\,, \qquad \mathbb{A} := \frac{\tfrac{b^2M}{a} + 2 \chi \mathbb{X}^5}{\left(\mathbb{X}+\tfrac{M}{2a}\right)^5}\,,
\end{equation}
which means that the energy density is always positive for $\chi \ge 0$. Moreover,  let us define 
\begin{equation}\label{M6}
\mathbb{B} := \frac{\mathbb{X} + \tfrac{M}{2a}}{\mathbb{X} - \tfrac{M}{2a}}\,, \qquad \mathbb{C} := \frac{\tfrac{b^2M^2}{a^2} - 4 \chi \mathbb{X}^6}{\left(\mathbb{X}+\tfrac{M}{2a}\right)^6}\,;
\end{equation}
then, the pressure in this case is given by 
\begin{equation}\label{M7}
p = -\frac{3 H^2}{8 \pi} - \frac{\dot{H}}{4\pi} \,\mathbb{B}+ \frac{1}{8 \pi a^2} \mathbb{B}\,\mathbb{C}\,.
\end{equation}

For $M \to 0$, we must recover the standard FLRW field equations. This is indeed the case, as we find
\begin{equation}\label{M8}
\mu_{M \to 0} =  \frac{3}{8\pi}\left(H^2 + \frac{k}{a^2\,R_0^2}\right),
\end{equation}
\begin{equation}\label{M9}
\left(p + \frac{\mu}{3}\right)_{M \to 0} = -\frac{1}{4\pi}\,\frac{\ddot{a}}{a}\,.
\end{equation}
These same limiting values are obtained if $b \to \infty$. 

In the other limiting situation, it is simple to verify that we recover the Plummer sphere in the background Minkowski spacetime, namely, 
\begin{equation}\label{M10}
\mu(a=1,\chi=0) = \mu_{\rm P}\,, \qquad  p(a=1,\chi=0) = p_{\rm P}\,.
\end{equation}

The motion of the perfect fluid source in this spacetime is vorticity free due to spherical symmetry and shear free due to spatially isotropic coordinates. The relative rate of volume expansion $\theta$ in this case works out to be 
\begin{equation}\label{M11}
\theta = U^{\mu}{}_{; \mu} = (-g)^{-1/2}\frac{\partial}{\partial x^\mu}[(-g)^{1/2}U^\mu] = 3 \,\frac{\dot a}{a}= 3 H(t)\,.
\end{equation}
Moreover, when the Hubble parameter is not constant, $b \ne 0$ or $\chi \ne 0$, there is in general a pressure singularity at $2a \mathbb{X} = M$. Consider the Kretschmann invariant,
\begin{equation}\label{M12}
R_{\alpha\beta\gamma\delta}R^{\alpha\beta\gamma\delta} = C_{\alpha\beta\gamma\delta}C^{\alpha\beta\gamma\delta} + 2 R_{\alpha\beta} R^{\alpha\beta} -\frac{1}{3} R^2\,,
\end{equation}
which can be worked out for the flat ($k = 0$) McVittie-Plummer spacetime, and the result is
\begin{align}\label{M13}
\nonumber \frac{P^2}{12}\,R_{\alpha\beta\gamma\delta} R^{\alpha\beta\gamma\delta} = {}& 2\, P^2 H^4+2\,PQ H^2 \dot{H}+ Q^2 \dot{H}^2 + \frac{Ma^2b^2}{\mathfrak{R}^5Q^5}\,\left[P(3P-Q)H^2-2\,aQ\dot{H}\right] \\   
& -\frac{M^2 a^4 (r^2-b^2)}{\mathfrak{R}^8 Q^{10}}(a^2+4\,P^2) +\frac{M^2a^4r^4}{\mathfrak{R}^{10} Q^{12}}\,(4\,a^2P^2 + 4\,P^2Q^2 + a^2 Q^2)\,. 
\end{align}
Here, we have introduced
\begin{equation}\label{M15}
P := a -\phi\,, \qquad Q := a + \phi\,, \qquad \mathfrak{R} = (r^2 + b^2)^{1/2} \,,
\end{equation}
such that $\phi = M/(2\mathfrak{R})$. For $b = 0$, the Kretschmann scalar reduces to the flat McVittie case, namely
\begin{equation}\label{M16}
R_{\alpha\beta\gamma\delta} R^{\alpha\beta\gamma\delta} = 12\left(2\, H^4+2\,\frac{Q}{P}\, H^2 \dot{H}+ \frac{Q^2}{P^2}\, \dot{H}^2 + \frac{4M^2a^6}{r^6Q^{12}}\right)\,.  
\end{equation}
The Kretschmann scalar diverges whenever $P = 0$ or  $\phi = a$, a curvature singularity that will be studied in more detail in the next section.

Let us now consider two further limiting cases. The Kretschmann scalar reduces to the FLRW case in the limit $M \to 0$, namely, 
\begin{equation}\label{M17}
R_{\alpha\beta\gamma\delta} R^{\alpha\beta\gamma\delta}|_{\text{FLRW}}= 12\left(2H^4+2\,H^2\,\dot{H} + \dot{H}^2 \right)\,.
\end{equation}
Moreover, in case  $b \to 0$ and $a(t) \to1$, we have the Schwarzschild limit
\begin{equation}\label{M18}
R_{\alpha\beta\gamma\delta} R^{\alpha\beta\gamma\delta}|_{\text{Schwarzschild}}= \frac{48\:M^2\, r^6}{(r + \tfrac{M}{2})^{12}}\,
\end{equation}
in Schwarzschild isotropic radial coordinate $r$. In terms of the standard Schwarzschild radial coordinate $r_{\rm Sch}$, 
\begin{equation}\label{M19}
r_{\rm Sch} = r \left( 1 + \frac{M}{2r}\right)^2\,, 
\end{equation}
Eq.~\eqref{M18} reduces to the familiar form $48 M^2/r_{\rm Sch}^6$, as expected. 

The Weyl curvature $C_{\mu \nu \rho \sigma}$ vanishes for FLRW metrics as they are conformally flat. The Weyl invariant $C_{\mu \nu \rho \sigma} C^{\mu \nu \rho \sigma}$ for the flat McVittie-Plummer spacetime thus turns out to be proportional to $M^2$ and is  given by 
\begin{equation}\label{M20}
C_{\alpha\beta\gamma\delta}C^{\alpha\beta\gamma\delta}=\frac{48\,M^2 r^4\,a^6}{\mathfrak{R}^{10}\,Q^{12}}\,.
\end{equation}

We now turn to the investigation of some of the main properties of the McVittie-Plummer spacetime. 

% @@@@@@@@@@@@@@@@@@@@@
% @@@@@@@@@@@@@@@@@@@@@
% @@@@@@@@@@@@@@@@@@@@@

\subsection{Singularity}

In the expression for pressure~\eqref{M7}, there is a physical singularity when $\mathbb{B}$ diverges at 
\begin{equation}\label{S1}
\mathbb{X}_{Singularity} = \frac{M}{2a}\,, \qquad a(t) = \phi\,K^{1/2}(r)\,, 
\end{equation}
which is therefore a spacetime singularity of the solution.  Let us note that this singularity is such that $(-g)^{1/2} = 0$; that is, the singularity corresponds, in effect, to the big bang singularity in this case.  At this singularity,  the energy density is in general finite and is given by
\begin{equation}\label{S2}
\mu_{Singularity} = \frac{3 H^2}{8 \pi} + \frac{3 \chi}{64 \pi\,a^2}+ \frac{3b^2\,a^2}{4\pi\,M^4}\,.
\end{equation}
If $M = 0$, then Eq.~\eqref{S1} implies $a = 0$ at the singularity and the energy density diverges as well, just as in the big bang singularity.  To our knowledge, this is the only singularity of the McVittie-Plummer spacetime for $b > 0$ and occurs at 
\begin{equation}\label{S3}
r_{Singularity} = \left[\frac{M^2 -4 a^2(t)\,b^2}{4 a^2(t) - \chi\, M^2}\right]^{1/2}\,.
\end{equation}
 For the flat $\chi = 0$ case, there is a singularity if $a(t) \le M/(2b)$. 

Consider the hypersurface $\mathcal{S}$ defined by 
 \begin{equation}\label{S4}
\mathcal{S}(t, r) := \mathbb{X} - \frac{M}{2a} - C_0\,, 
\end{equation}
where $C_0$ is a constant. The 4-vector normal to this hypersurface $n_\mu = \partial_\mu \mathcal{S}$ is such that 
\begin{equation}\label{S5}
n_\mu n^\mu = - \frac{M^2 H^2}{4a^2} \mathbb{B}^2 + \frac{K^2\,\mathbb{X}^4}{a^2}\, \left(\mathbb{X} + \tfrac{M}{2a}\right)^{-4} \left(\frac{d\mathbb{X}}{dr}\right)^2\,.
\end{equation}
If $H \ne 0$, the spacetime is dynamic and for $C_0 = 0$, the singular hypersurface where $\mathbb{B}$ diverges is spacelike. On the other hand, if $H = 0$, then we have a spherically symmetric static cosmological model with a singularity at $C_0 = 0$ that is a timelike or possibly a null hypersurface; in fact,  the latter possibility occurs for $k = 1$ and $b = 2 R_0$.  That is, we note that 
\begin{equation}\label{S6}
\frac{d\mathbb{X}}{dr} = \frac{\mathbb{X}\,r(1-\chi b^2)}{(r^2+b^2)(\chi\,r^2 + 1)}\,, 
\end{equation}
which vanishes when $k = 1$ and $b = 2R_0$. 

The investigation of timelike and null geodesics as well as the horizon structure of the McVittie-Plummer spacetime is beyond the scope of the present work. 

%@@@@@@@@@@@@@@
%@@@@@@@@@@@@@@

\subsection{Weak Energy Condition}

The perfect-fluid source has energy density and pressure; these  material properties must be related through some kind of equation of state. A simple relation like $p = w \mu$ with constant $w$ is too simple to work in this case. The solution has a proper physical interpretation only if a physically reasonable equation of state exists; otherwise, one must think of other possibilities for $T_{\mu \nu}$. In the absence of a clear connection between $p$ and $\mu$, we must at least ensure that the energy-momentum tensor satisfies the weak energy condition~\cite{HE}. Let $\Upsilon := T_{\mu \nu} W^\mu W^\nu$ for an arbitrary timelike (or null) vector $W^\alpha$  such that $W^\mu W_\mu \le 0$; then, the \emph{weak energy condition} is satisfied provided $\Upsilon \ge 0$. For our energy-momentum tensor~\eqref{M2}, we have 
\begin{equation}\label{N1}
\Upsilon  = (\mu + p)(U_\alpha W^\alpha)^2 + p W_\beta\,W^\beta\,.
\end{equation}

For an arbitrary null vector field $W_\alpha\,W^\alpha = 0$, we must clearly have $\mu + p \ge 0$. On the other hand, for an arbitrary timelike vector field $W_\alpha\,W^\alpha < 0$, let us write
\begin{equation}\label{N2}
\Upsilon  = (\mu + p)\left[ (U_\alpha W^\alpha)^2 +  W_\beta\,W^\beta\right] - \mu W_\beta\,W^\beta\,.
\end{equation}
We recall from Eq.~\eqref{M2} the explicit relation for $U^\mu$ and then employ the diagonal form of the metric to show that 
\begin{equation}\label{N3}
 (U_\alpha W^\alpha)^2 +  W_\beta\,W^\beta =  g_{ij} W^i W^j \ge 0\,.
\end{equation}
Hence, in general $\Upsilon \ge 0$ in Eq.~\eqref{N2} provided $\mu \ge 0$. 

The weak energy condition is thus satisfied in the McVittie-Plummer spacetime for $\mu \ge 0$ and $\mu + p \ge 0$. More explicitly, the condition for the energy density is clearly satisfied for $\chi \ge 0$. For $\mu + p$, let us note that 
\begin{equation}\label{N4}
\mu + p = \frac{\mathcal{N}}{\mathcal{D}}\,,
\end{equation}
where 
\begin{equation}\label{N5}
\mathcal{N} = 3 \left(\mathbb{X} - \tfrac{M}{2a}\right)\left(\tfrac{b^2M}{a} + 2 \chi \mathbb{X}^5\right) - a^2 \dot{H}\,\left(\mathbb{X} + \tfrac{M}{2a}\right)^6+ \frac{b^2M^2}{2a^2} -2\chi \mathbb{X}^6\,,
\end{equation}
\begin{equation}\label{N6}
\mathcal{D} = 4 \pi a^2 \left(\mathbb{X} - \tfrac{M}{2a}\right)\left(\mathbb{X} + \tfrac{M}{2a}\right)^5\,.
\end{equation}
For the rest of this discussion, we assume the flat case $\chi = 0$. Now the weak energy condition is satisfied beyond the big bang singularity (i.e. $\mathbb{X} > \tfrac{M}{2a}$), if $-\dot{H} = (1+q) H^2  \ge 0$; that is, the rate of expansion of the universe should either be constant or decrease forever.

%@@@@@@@@@@@
%@@@@@@@@@@@
%@@@@@@@@@@@

\section{Preferred Observers}

Consider the flat McVittie-Plummer metric
\begin{equation}\label{J1}
 ds^2 = - \frac{P^2}{Q^2}\,dt^2 + \frac{Q^4}{a^2(t)}\,(dr^2 + r^2 d\Omega^2)\,,
\end{equation} 
where
\begin{equation}\label{J2}
 d\Omega^2 =  d\vartheta^2 + \sin^2 \vartheta \, d\varphi^2\,
\end{equation} 
in spherical polar coordinates. 
Here, we have set $k = 0$ for the sake of simplicity; moreover, $x^\mu = (t, r, \vartheta , \varphi)$ and  $(P, Q) = (a - \phi, a +\phi)$ as in Eq.~\eqref{M15}, where $\phi$ is given by Eq.~\eqref{R2}. The preferred observers are the comoving observers of standard cosmology that stay at rest in space and have 4-velocity vectors
\begin{equation}\label{J2a}
U^\mu = dx^\mu /d\tau = (Q/P) \,\delta^\mu_0\,.
\end{equation} 
These observers have translational acceleration $\mathcal{A}^\mu$ given in $(t, r, \vartheta, \varphi)$ coordinates by
\begin{equation}\label{J3}
\frac{DU^\mu }{d\tau} = \mathcal{A}^\mu\,, \qquad \mathcal{A}^\mu = \frac{Ma^3 r}{PQ^5 (r^2+b^2)^{3/2}}\, (0, 1, 0, 0)\,,
\end{equation} 
where we have used the connection coefficients given in Appendix A. 

More generally, the preferred observer carries an adapted orthonormal tetrad frame $e^{\mu}{}_{\hat \alpha}$,
\begin{equation}\label{J4}
e^{\mu}{}_{\hat 0} := U^\mu = \frac{Q}{P} \,\delta^\mu_0\,,  \quad e^{\mu}{}_{\hat 1} := \frac{a}{Q^2}\, \delta^\mu_{1}\,,\quad e^{\mu}{}_{\hat 2} := \frac{a}{Q^2\,r}\, \delta^\mu_{2}\,, \quad e^{\mu}{}_{\hat 3} := \frac{a}{Q^2\,r \sin\vartheta}\, \delta^\mu_{3}\,.
\end{equation} 
The local spatial frame of the observer has unit axes that point along the spatial coordinate directions. We define the acceleration tensor $\Phi_{\hat \alpha \hat \beta}$,
\begin{equation}\label{J5}
\frac{D e^{\mu}{}_{\hat \alpha}}{d\tau} = \Phi_{\hat \alpha}{}^{\hat \beta} \, e^{\mu}{}_{\hat \beta}\,,
\end{equation} 
whose elements correspond to the translational and rotational accelerations of the observer. It follows from the orthonormality relation
\begin{equation}\label{J6}
g_{\mu \nu}\,e^{\mu}{}_{\hat \alpha}\,e^{\nu}{}_{\hat \beta} = \eta_{\hat \alpha \hat \beta}\,
\end{equation} 
that the acceleration tensor is antisymmetric, namely, $\Phi_{\hat \alpha \hat \beta} = - \Phi_{\hat \beta \hat \alpha}$. We find in the present case that
\begin{equation}\label{J7}
\Phi_{\hat 0 \hat 1} = - \Phi_{\hat 1 \hat 0}= \mathfrak{g}_{\hat 1} = \frac{M r\,a^2}{PQ^3(r^2+b^2)^{3/2}}\,
\end{equation} 
and all others vanish. That is, the local spatial frame of the preferred observer is nonrotating (i.e.  Fermi-Walker transported) along its world line.  Note that $\mathfrak{g}_{\hat 1} = \mathcal{A} \cdot e_{\hat 1}$ is positive, which means that for the observer to stay fixed in space, this quantity balances the attraction of gravity. In this connection, $\mathfrak{g}_{\hat 1} \to \infty$ for $\phi \to a$, which means that the attraction of gravity is infinite at the singularity, while $\mathfrak{g}_{\hat 1} \to M/(a^2r^2)$ for $r \to \infty$, which is the expected Newtonian result and indicates that $\mathbb{R}:= a(t)\,r$ is an appropriate Newtonian radial distance.  

Changing the radial coordinate $r$ in metric~\eqref{J1} to $\mathbb{R} = a(t)\, r$, we find
\begin{equation}\label{J8}
 ds^2 = - \frac{\mathbb{P}^2}{\mathbb{Q}^2}\,dt^2 + \frac{\mathbb{Q}^4}{16[\mathbb{R}^2 + a^2(t)b^2]^2}\,[(d\mathbb{R} - H\mathbb{R} dt)^2 + \mathbb{R}^2\,d\Omega^2]\,,
\end{equation} 
where
\begin{equation}\label{J9}
a(t) dr =  d\mathbb{R} - H\mathbb{R}\, dt\,
\end{equation} 
and
\begin{equation}\label{J10}
\mathbb{P} := 2\,[\mathbb{R}^2 + a^2(t)b^2]^{1/2} - M\,,\qquad \mathbb{Q} := 2\,[\mathbb{R}^2 + a^2(t)b^2]^{1/2} + M\,.
\end{equation} 
The singularity occurs at $\mathbb{P} = 0$. 

To express metric~\eqref{J8} in a more familiar form, let us introduce a radial coordinate that is a simple generalization of $r_{\rm Sch}$, namely, $\mathfrak{r}$ given by
\begin{equation}\label{J11}
\mathfrak{r} := \frac{1}{4}\,\frac{ \mathbb{Q}^2}{[\mathbb{R}^2 + a^2(t)b^2]^{1/2}}\,;
\end{equation} 
then, we find
\begin{equation}\label{J12}
[\mathbb{R}^2 + a^2(t)b^2]^{1/2} = \frac{1}{4}\,\mathfrak{r}\left[ 1 + \left(1 - \frac{2M}{\mathfrak{r}}\right)^{1/2}\right]^2\,, \qquad  \frac{\mathbb{P}^2}{\mathbb{Q}^2} = 1 - \frac{2M}{\mathfrak{r}}\,. 
\end{equation}
After some algebra, metric~\eqref{J8} takes the form
\begin{equation}\label{J13}
 ds^2 = - \left(1 - \frac{2M}{\mathfrak{r}} - H^2 \eta^{-2}\mathfrak{r}^2 \right)\,dt^2 + \frac{2 \eta^{-2}\mathfrak{r}H}{ \left(1 - \frac{2M}{\mathfrak{r}}\right)^{1/2}}\, dt d\mathfrak{r} + \eta^{-2} \frac{d\mathfrak{r}^2}{1 - \frac{2M}{\mathfrak{r}}}+ \eta^2\,\mathfrak{r}^2d\Omega^2\,,
\end{equation} 
where $\eta$, $0 < \eta \le 1$, is given by
\begin{equation}\label{J14}
\eta :=  \frac{\mathbb{R}}{[\mathbb{R}^2 + a^2(t)b^2]^{1/2}}\,, \qquad \frac{ a^2(t) b^2}{1-\eta^2} =\frac{1}{4}\, \{\mathfrak{r} - M + [\mathfrak{r}(\mathfrak{r}-2M)]^{1/2}\}^2\,.
\end{equation} 
For $b = 0$, $\eta = 1$ and metric~\eqref{J13} reduces to the flat ($k=0$) McVittie metric, which for  $H = {\rm constant}$ further reduces via a simple coordinate transformation to the Schwarzschild-de Sitter metric of Kottler's spacetime~\cite{Kottler}.

 % @@@@@@@@@@@@@@@@@@@@@
% @@@@@@@@@@@@@@@@@@@@@
% @@@@@@@@@@@@@@@@@@@@@

\section{Mass-Energy Function}

Imagine an arbitrary test observer in a cosmological background. The observer can establish a quasi-inertial Fermi normal coordinate system in its spacetime neighborhood. The deviation of the resulting metric from Minkowski spacetime in the limited cylindrical domain along the world line of the observer has to do with the acceleration tensor of the observer as well as the local Riemannian curvature caused by the cosmic distribution of matter throughout spacetime.  The nature of this distribution is admittedly rather complex, as evidenced by the cosmic web. On the other hand, if one imagines that the structure of the cosmic web is replaced by a uniform distribution of matter that ensures spatial homogeneity and isotropy, then the geometry of spacetime is given by a FLRW cosmological model.  In this simplified framework, a comoving observer follows a geodesic with its proper time equal to cosmic time $t$; moreover, its adapted tetrad frame $\lambda^{\mu}{}_{\hat \alpha}$, given by Eq.~\eqref{H3}, is parallel transported along its world line. A geodesic  (Fermi) normal coordinate system can be based on this adapted tetrad frame. The spatial homogeneity of the FLRW background may make it possible to introduce exact Fermi normal coordinates once the temporal dependence of the scale factor $a(t)$ is explicitly known~\cite{Chicone:2005vn}. In general, however, a general perturbation scheme may be employed to determine the approximate form of the metric. 

To illustrate this perturbative approach, imagine an event at some proper time $t$ along the geodesic world line of the reference observer and all spacelike geodesics that start at this event and are normal to the world line of the reference observer. These form a local spacelike hypersurface. Let $x^\mu$ be an event on this local hypersurface that can be connected to the reference observer's geodesic world line at event $t$ by a unique spacelike geodesic segment  of proper length $\sigma$. Then, we assign to event $x^\mu$ Fermi coordinates $(T, \mathbf{X})$ such that $T = t$ and $X^i = \sigma \,\xi^\mu \lambda_{\mu}{}^{\hat i}$, where $\xi^\mu$ is the unit spacelike  vector tangent to the unique geodesic segment at event $t$. Let us note that the reference observer is forever fixed at the spatial origin of the local quasi-inertial Fermi coordinate system. A detailed treatment within the framework of standard cosmological models is contained in~\cite{Mashhoon:2007qm, Mashhoon}, where the approximate Fermi metric valid to second order in the distance away from the observer is described. It is convenient to introduce spherical polar coordinates in the Fermi system, $\mathbf{X} \mapsto (\rho, \Theta, \Phi)$, where
\begin{equation}\label{F1}
X^1  = \rho \sin \Theta\,\cos \Phi\,, \qquad X^2 = \rho \sin \Theta\,\sin \Phi\,, \qquad X^3 = \rho \cos \Theta\,.
\end{equation} 
Then, the Fermi metric for the standard comoving reference observer is given by
\begin{equation}\label{F2}
 ds^2 = - [1 + \Pi_1 (T)\, \rho^2]\,dT^2 + d\rho^2 + \rho^2\,[ 1 - \tfrac{1}{3} \Pi_2(T)\,\rho^2]\, (d\Theta^2 + \sin^2 \Theta \, d\Phi^2)\,,
\end{equation} 
where the gravitational potentials are valid to second order in the radial Fermi coordinate $\rho$, which should be small in comparison with the local radius of curvature of spacetime to ensure the admissibility of  the Fermi coordinate system~\cite{Chicone:2005vn}. Here, $\Pi_1$ and $\Pi_2$ are as defined in Eq.~\eqref{H8}. It is clear that any local physical system within the domain of admissibility of the Fermi coordinates will experience the effects of the spacetime curvature of the rest of the cosmos. That is, the spacetime curvature may be pointwise ignored in accordance with Einstein's principle of equivalence, but any finite system will be affected by it. 

On the other hand, the simplicity of McVittie and related solutions of general relativity rests on the fact that what is embedded in the FLRW spacetime is not directly affected by the cosmic background; for instance, no accretion of matter takes place. Nevertheless, Cadoni et al.~\cite{Cadoni:2023lqe, Cadoni:2024jxy} have suggested that the influence of the cosmic background may show up in the invariant physical quantities associated with extended spacetime regions such as the Misner-Sharp mass-energy function. The purpose of this section is to show that this is indeed the case for the McVittie-Plummer spacetime. 

The mass-energy function is given by Eq.~\eqref{R11}. We calculate it in this specific case which results  in three terms, the first two are sourced by the Plummer sphere and vanish when $M = 0$, while the last term is mainly due to the cosmic background. To see this in detail, let us write the McVittie-Plummer  metric~\eqref{M1} in terms of spherical polar coordinates in the form
\begin{equation}\label{F3}
 ds^2 = - \frac{1}{\mathbb{B}^2} \,dt^2 + \mathbb{D}^2\,(dr^2 + r^2 d\Omega^2)\,,
\end{equation}
where
\begin{equation}\label{F4}
 \mathbb{D} :=  \frac{a(t)}{r^2+b^2}\,\left(\mathbb{X}+\tfrac{M}{2a}\right)^2\,.
\end{equation}
Let $\dot{\mathbb{D}} := \partial \mathbb{D}/\partial t$ and $\mathbb{D}' := \partial \mathbb{D}/\partial r$; then, for the McVittie-Plummer spacetime the Misner-Sharp mass, $m_{\rm MP}$, is given by
\begin{equation}\label{F5}
1 - \frac{2m_{\rm MP}}{r\mathbb{D}} = \left(1+ \frac{r\mathbb{D}'}{\mathbb{D}}\right)^2 - (r\mathbb{B}\,\dot{\mathbb{D}})^2\,,
\end{equation}
where 
\begin{equation}\label{F6}
r\mathbb{B}\,\dot{\mathbb{D}} = \frac{r\dot{a}}{r^2+b^2}\left(\mathbb{X}+\tfrac{M}{2a}\right)^2\,,
\end{equation}
\begin{equation}\label{F7}
\frac{\mathbb{D}'}{\mathbb{D}} = -\frac{2r}{r^2+b^2} +  \frac{2}{\mathbb{X}+\tfrac{M}{2a}}\,\frac{d\mathbb{X}}{dr}\,
\end{equation}
and $d\mathbb{X}/dr$ is given by Eq.~\eqref{S6}. Let us recall that $\mathbb{X}^2 = (r^2+b^2)/(\chi\, r^2 + 1)$; therefore, Eq.~\eqref{F7} can be written as
\begin{equation}\label{F8}
\frac{\mathbb{D}'}{\mathbb{D}} = -\frac{2r}{r^2+b^2}\,\left(\mathbb{X}+\tfrac{M}{2a}\right)^{-1} \left(\chi\,\mathbb{X}^3+\tfrac{M}{2a}\right)\,.
\end{equation}
After some algebra, we finally arrive at the Misner-Sharp mass for the McVittie-Plummer spacetime
\begin{equation}\label{F9}
m_{\rm MP} (t, r) = \frac{2ar^3}{(r^2+b^2)^3}\left[\left(\mathbb{X}^3+\tfrac{Mb^2}{2a}\right)\left(\chi\,\mathbb{X}^3+\tfrac{M}{2a}\right) + \frac{1}{4} a^2 H^2\,\left(\mathbb{X}+\tfrac{M}{2a}\right)^{6}\right]\,.
\end{equation}

It is interesting to write the final result in the form
\begin{equation}\label{F10}
m_{\rm MP} (t, r) =  \frac{Mr^3\mathbb{X}^3(1+\chi\,b^2)}{(r^2+b^2)^3} + \frac{M^2b^2r^3}{2a(r^2+b^2)^3}+ \frac{2ar^3}{(r^2+b^2)^3} \left[\chi\,\mathbb{X}^6 + \frac{1}{4} a^2 H^2\,\left(\mathbb{X}+\tfrac{M}{2a}\right)^{6}\right]\,.
\end{equation}
When the Plummer radius vanishes, i.e. $b=0$, we have the McVittie limit, which for the flat case ($\chi = 0$) simplifies to
\begin{equation}\label{F11}
m_{\rm M} (\chi = 0) = m_{\rm MP}(b=0, \chi = 0) = M +\frac{1}{2} \,r^3 a^3\,(1+\frac{M}{2ra})^6\,H^2.
\end{equation}
That is, the astrophysical point source in the first term is simply unaffected by the expansion of the universe, while the second term that involves the mass-energy due to the cosmic background does contain a contribution from the point mass. For $M = 0$, we are simply left with the contribution of the FLRW background, namely,  $m_{\rm FLRW} (k = 0) = (4\pi/3)(ra)^3 \mu_{\rm FLRW} (k = 0) $, where $\mu_{\rm FLRW} (k = 0) = 3H^2/(8 \pi)$. 

Let us now return to the general situation in Eq.~\eqref{F10} and note that the first term in this expression is of the form
\begin{equation}\label{F12}
\frac{Mr^3}{(r^2+b^2)^{3/2}}\,\frac{1+\chi\,b^2}{(1 + \chi\,r^2)^{3/2}} = M(<r) \,\frac{1+\chi\,b^2}{(1 + \chi\,r^2)^{3/2}}\,,
\end{equation}
where $M(<r)$ is the purely Newtonian contribution due to the presence of the embedded Plummer sphere,  while the second term in Eq.~\eqref{F10} is a time-dependent general relativistic correction, namely, 
\begin{equation}\label{F13}
M(<r) \,\frac{GMb^2}{2c^2(r^2+b^2)^{3/2}a(t)}\,. 
\end{equation}

To simplify matters, we write  Eq.~\eqref{F10} for the flat ($k=0$ ) case, where $\chi = 0$ and  $\mathbb{X}= (r^2+b^2)^{1/2}$, 
\begin{equation}\label{F14}
m_{\rm MP}(\chi = 0) = M (< r) \,\left[1 + \frac{GMb^2}{2c^2(r^2+b^2)^{3/2}\,a(t)}\right] + \frac{c^2}{2G}\, r^3 a^3 H^2\,\left[ 1 + \frac{GM}{2c^2a (r^2+b^2)^{1/2}} \right]^6\,,
\end{equation}
and note that the last term proportional to $H^2$ is the mass-energy of the flat FLRW background augmented by the relativistic contribution of the Plummer sphere; that is, $\mu_{\rm FLRW} (k = 0) = 3 c^2 H^2/(8 \pi G)$ and 
\begin{equation}\label{F15}
 \left(\frac{3c^2H^2}{8 \pi G}\right) \frac{4 \pi}{3}\,(r a)^3  = \frac{c^2}{2G}\, r^3 a^3 H^2\,. 
\end{equation}
For $a=1$, $H = 0$ and Eq.~\eqref{F14} reduces to Eq.~\eqref{R12}, which represents the Misner-Sharp mass-energy function for the relativistic Plummer sphere.
 
To sum up, the influence of the cosmic expansion on the astrophysical system shows up in the second term in our general result~\eqref{F10}. This term is independent of $\chi$, represents the gravitational coupling of the Plummer sphere with the expanding background FLRW model and for $r\to \infty$ vanishes as $r^{-3}$. It is part of the interplay between the cosmological expansion and the extended massive object; in fact, the second term depends upon cosmic time as $1/a(t)$ and upon the finite size of the embedded source as it is  proportional to the square of the Plummer radius. This coupling decreases in time as the universe expands. 

%%%%%%%%%%
%%%%%%%%%%

\section{Observational Consequences}

To examine possible observational consequences of  the McVittie-Plummer model, we imagine throughout this section that the Plummer sphere is embedded in the current benchmark model of $\Lambda$CDM cosmology~\cite{Dodelson:2003ft,Weinberg:2008zzc,Amendola:2015ksp,Baumann:2022mni}. To characterize the extent of the overdensity in our model, we define $R_{200}$ as radius of a sphere within which the overdensity is $200$ times the background cosmic matter density $\bar{\mu}$; that is, 
\begin{equation}\label{h1}
M = 200\, (\tfrac{4}{3}\pi R_{200}^3)\, \bar{\mu}\,, \qquad GM = 100\,H_0^2\, R^3_{200} \Omega_{m,0} (1+z)^3\,,
\end{equation}
where $\Omega_{m,0} = \bar{\mu} / \bar{\mu}_c$ is the density of background matter at the present time normalized to the critical density $\bar{\mu}_c = 3\,H_0^2 / (8\pi G)$.  Hence, $R_{200}$  is an approximate indicator of the size of an overdense region~\cite{Gunn:1972sv} and, by definition,  decreases with increasing cosmological redshift as $(1+z)^{-1}$. 

To illustrate the integration of the overdensity into the background, we turn to graphical representations of $\mu$, $p$ and $m_{\rm MP}$. \\

%%%%%%%%%
%%%%%%%%%

\subsection{Plots of energy density, pressure and Misner-Sharp mass-energy function}

\begin{figure}
	\centering
	\includegraphics[scale=0.7]{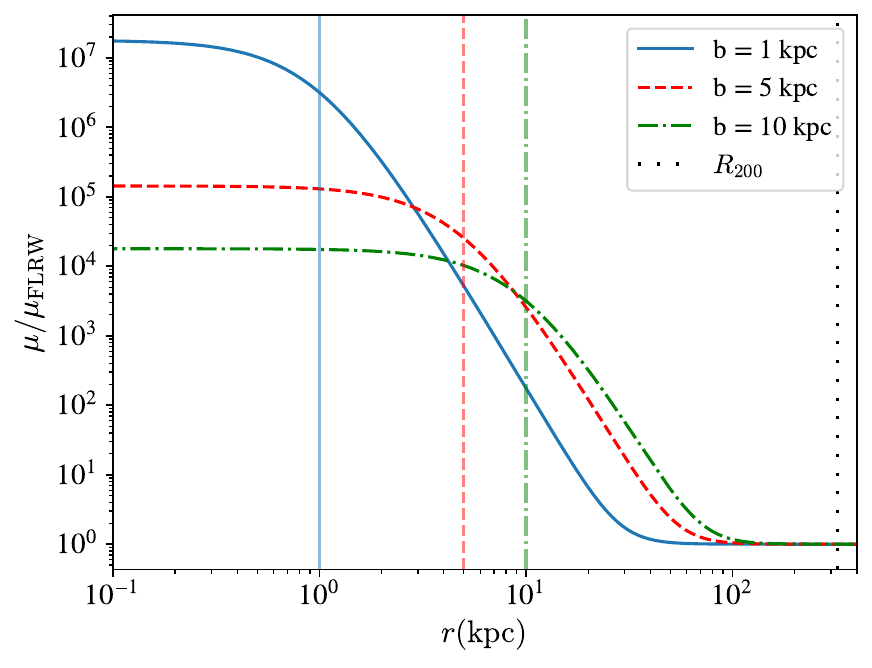}
	\caption{Plot of $\mu / \mu_{\rm FLRW}$ at the present epoch versus the radial coordinate $r$ for three Plummer radii: $b = 1$ kpc (solid blue), $b = 5$ kpc (red dash) and $b = 10$ kpc (green dot-dash). A corresponding vertical line indicates the magnitude of the Plummer radius in each case. The black dotted line shows the magnitude of $R_{200}$.}\label{fig:1}
\end{figure}

For the plots that follow, we assume that  $M = 10^{10}M_{\odot}$ is the mass of the Plummer sphere.   The energy density of the McVittie-Plummer model for the flat case ($\chi = 0$) follows from Eq.~\eqref{M5}; that is, 
\begin{equation}\label{h2}
\mu =  \frac{3c^2H^2}{8 \pi G}+ a^{-3} \frac{Mc^2}{(4 \pi/3) b^3}\,\left(1 + \frac{r^2}{b^2}\right)^{-5/2}\left( 1 + \frac{GM}{2c^2a\sqrt{r^2+b^2}}\right)^{-5}\,. 
\end{equation}
In Figure~\ref{fig:1}, we plot $\mu / \mu_{\rm FLRW}$ at the present epoch versus the radial coordinate $r$, where $\mu_{\rm FLRW} = 3c^2H^2 /8 \pi G$ for the flat case under consideration. The plot is for three Plummer radii, namely, $b= (1,5,10)$ kpc, indicated by vertical lines. As is clear from the analytical formula~\eqref{h2}, there is almost a plateau in small radii up to the Plummer radius $b$. Over larger scales $r>100$ kpc,  the density asymptotically approaches the cosmological background density. 
From $a^{-1} = 1+z$  and Eq.~\eqref{h2}, it is clear that $\mu / \mu_{\rm FLRW}$ is expected to vary with the cosmological redshift $z$ mainly as $(1+z)^3/H^2$.  If the expansion of the universe is dominated by the cosmological constant, $H$ is nearly constant; hence,  $\mu / \mu_{\rm FLRW}$ should increase as $(1+z)^3$. In the case under consideration here, we have verified this result for $0 \le z \le1$.

\begin{figure}
	\centering
	\includegraphics[scale=0.7]{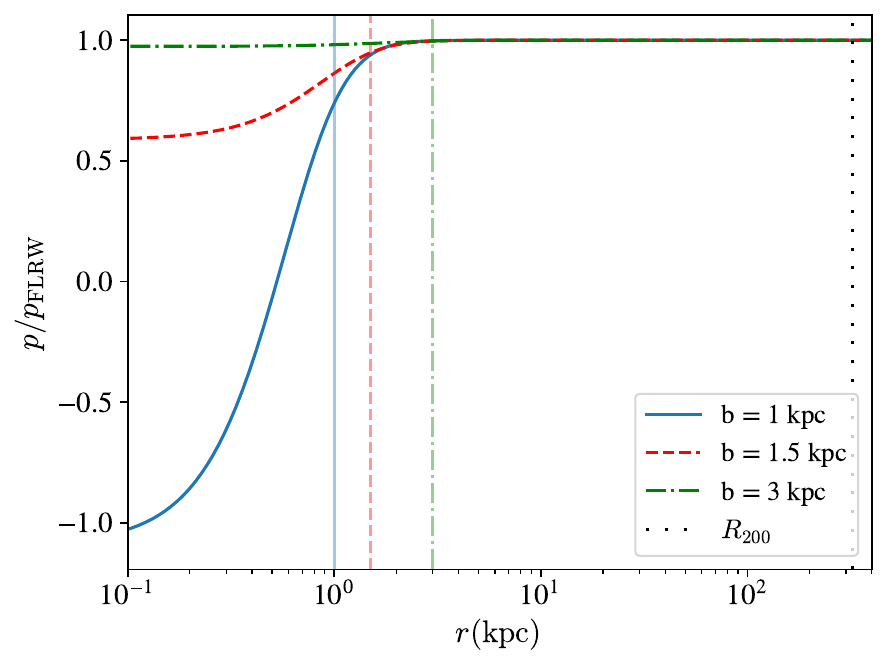}
	\caption{Plot of $p/p_{\rm{FLRW}}$ at the present epoch versus the radial coordinate $r$ for three Plummer radii: $b = 1$ kpc (solid blue), $b = 1.5$ kpc ( red dash) and $b = 3$ kpc (green dot-dash). A corresponding vertical line indicates the magnitude of the Plummer radius in each case. The black dotted line shows the magnitude of $R_{200}$.}\label{fig:2}
\end{figure}

The pressure of the McVittie-Plummer model for the flat case ($\chi = 0$) is given by Eq.~\eqref{M7}; in this case, the analytic expression is more complicated and the pressure diverges at the spacetime singularity of this model. In Figure~\ref{fig:2}, we plot the ratio $p/p_{\rm{FLRW}}$ for the overdense region at the present epoch versus the radial coordinate $r$ for $b=(1,\:1.5,\:3)$ kpc. Here, $p_{\rm FLRW} = - \tfrac{3c^2}{8\pi G}H^2 - \tfrac{c^2}{4\pi G}\dot{H} $ is the background FLRW pressure and its magnitude is negative  in the current epoch of the flat $\Lambda$CDM model. In Figure~\ref{fig:2}, as $b$ increases, the pressure in the overdense region increases as well, while for radii larger than $\approx 1$ kpc, the pressure rapidly approaches its background value.   
Regarding the redshift dependence of the pressure, we have verified that  $p/p_{\rm{FLRW}}$ decreases when $z$ increases  within the range $0 \le z \le1$.

\begin{figure}
	\centering
	\includegraphics[scale=0.7]{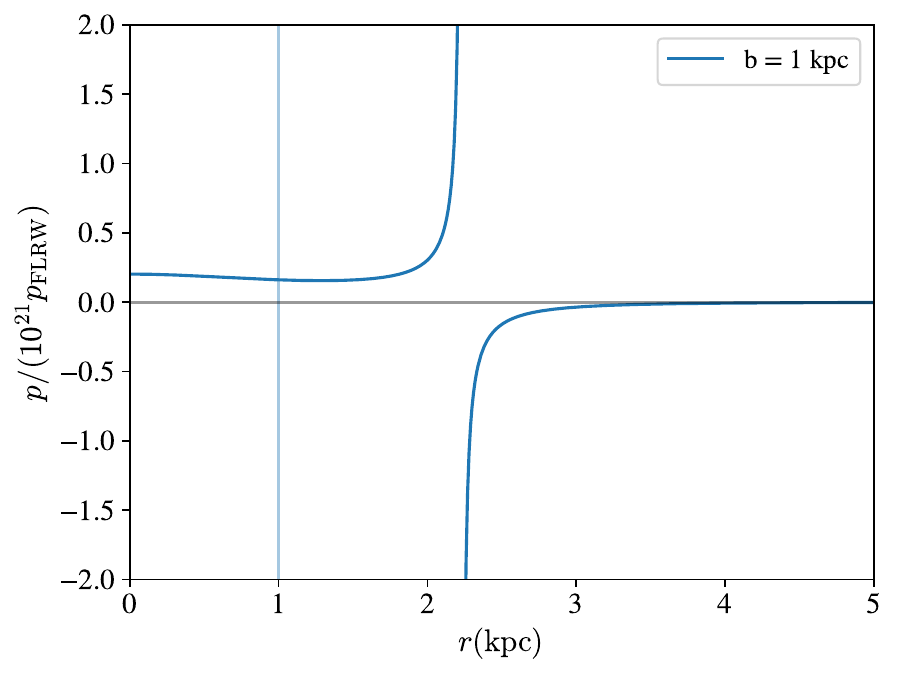}
	\caption{Plot of $p/(10^{21}\,p_{\rm{FLRW}})$ versus $r$ for Plummer radius $b = 1$ kpc (solid blue) at redshift $z=10^7$. The vertical line indicates the magnitude of the Plummer radius. The divergence of the pressure occurs at $r \approx 2$ kpc.}\label{fig:3}
\end{figure}

An important aspect of the pressure in the McVittie-Plummer model is its divergence, which we explore in Figure~\ref{fig:3}. The singularity occurs at a value of the radial coordinate $r$ and scale factor  $a$ such that $GM = 2c^2a (r^2 + b^2)^{1/2}$. With $M = 10^{10}M_{\odot}$, we have $GM/c^2 \approx 1.5 \times 10^{15}$ cm. Moreover, with $a = (1+z)^{-1}$, redshift $z = 10^7$ and $b=1$ kpc, we find that the singularity occurs at a radius $ r \approx 2$ kpc, in agreement with Figure~\ref{fig:3}. 

\begin{figure}
	\centering
	\includegraphics[scale=0.7]{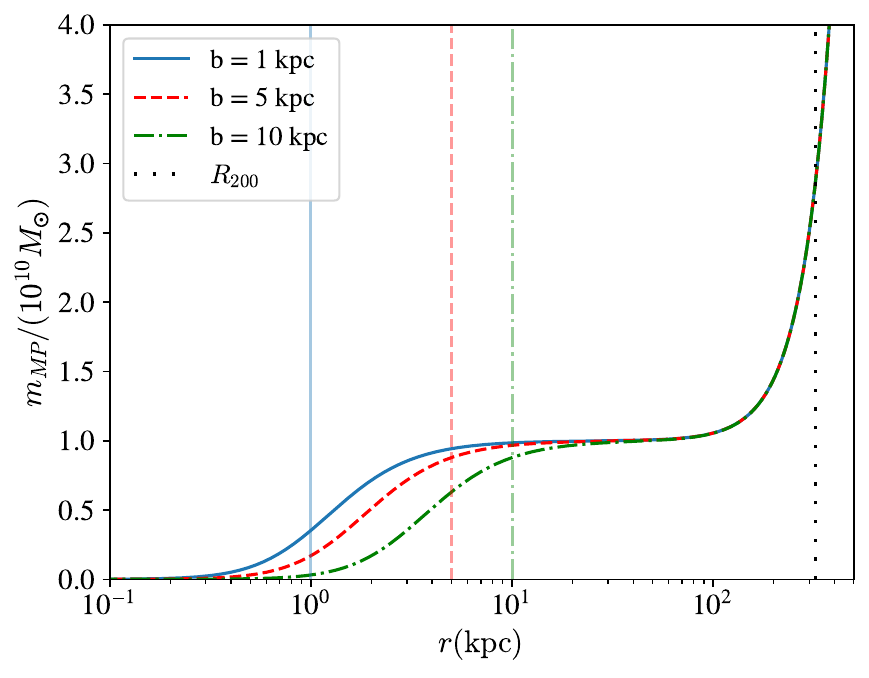}
	\caption{Plot of $m_{\rm{MP}}/M$ at the present epoch versus the radial coordinate $r$ for three Plummer radii: $b = 1$ kpc (solid blue), $b = 5$ kpc (red dash) and $b = 10$ kpc (green dot-dash). A corresponding vertical line indicates the magnitude of the Plummer radius in each case. The black dotted line shows the magnitude of $R_{200}$.}\label{fig:4}
\end{figure}

Regarding the Misner-Sharp mass-energy function $m_{\rm{MP}}(t, r)$,  we plot $m_{\rm{MP}}/M$ in Figure~\ref{fig:4} at the present epoch versus the radial coordinate $r$ for three different Plummer radii $b= (1,\:5,\:10)$ kpc. The Misner-Sharp mass $m_{\rm{MP}}$ starts from zero at $r=0$ and for $r > b$ eventually converges to the function $M[1 + H^2 a^3 r^3 /(2GM)]$. In the $\Lambda$CDM benchmark model, the $H^2 a^3$ factor is expected to decrease as the cosmological redshift increases.  We have verified this cosmological redshift  dependence for $b = 5$ kpc and $z$ in the range $0 \le z \le1$.

%%%%%%%
%%%%%%%

\subsection{Determination of Plummer Radius $b$}

The McVitte-Plummer spacetime represents a smooth spherically symmetric distribution of matter embedded in the FLRW universe. The overdensity could be due to a cluster of stars or the dark matter halo of a spherical dwarf galaxy.  It is a valuable quest to use observational data to constrain the Plummer radius $b$. The mass and distribution of clustered matter can be studied by gravitational lensing. For instance, in the thin lens approximation of strong lensing scenario, it is possible to measure the mass distribution projected along the line of sight; in this way, the average value of the Plummer radius can be estimated~\cite{Schneider:2006qyj}. We note in this connection that the lensing of the Plummer model has been studied in~\cite{Saha:2024axf}; moreover, lensing in the McVittie universe has been studied in~\cite{Piattella:2015xga, Piattella:2016nzt, Bessa:2022sdh}. The other observational method to constrain the density profile of a galaxy could be the corresponding rotation curve. The main contribution of the rotation curve for fixing the mass profile has to do with the outer regions of a galaxy~\cite{Persic:1995ru}, which is used to study the structure of the host dark matter halo of a galaxy~\cite{Dubinski:1991bm}. As the relativistic Plummer overdensity modifies the inner region, this method is not expected to be efficient for constraining the Plummer $b$ parameter~\cite{Saha:2024axf}. Finally, an interesting observational probe would be  the surface light profile~\cite{Noyola:2006mg}.  With a plausible mass-to-light ratio, one can compare the two-dimensional projected density profile to determine the Plummer radius. It is important to note that this projected mass is a function of the cosmological redshift due to the coupling of astrophysics and cosmology demonstrated in Section VI.

\section{DISCUSSION}

The McVittie-Plummer cosmological model describes the embedding of a spherical galaxy in a background FLRW cosmological model and is a simple generalization of McVittie model. The energy density and pressure of the perfect fluid source of the McVittie-Plummer gravitational field are determined and the conditions under which the weak energy condition is satisfied are specified. The Misner-Sharp mass-energy function $m (t, r)$  is calculated in this case and the mutual interaction between the extended astrophysical system and the expanding cosmic background within the Misner-Sharp mass is elucidated.  A brief comparison of the McVittie-Plummer model with the LTB model is contained in Appendix B. 

%@@@@@@@@@@@

%@@@@@@@@@@@@@@@@@@

\section*{ACKNOWLEDGMENTS}

We thank Masoud Molaei for helpful discussions.
SB is partially supported by the Abdus Salam International Center for Theoretical Physics (ICTP) under the regular associateship scheme.
Moreover, JT and SB are partially supported by the Sharif University of Technology Office of Vice President for Research under Grant No. G4010204.

\appendix

\section{Christoffel Symbols}

We list below the nonzero connection coefficients for  the flat ($k = 0$) McVittie-Plummer metric
\begin{equation}\label{A1}
 ds^2 = - \frac{P^2}{Q^2}\,dt^2 + \frac{Q^4}{a^2(t)}\,(dr^2 + r^2\, d\vartheta^2 + r^2 \sin^2 \vartheta\, d\varphi^2)\,
\end{equation} 
in spherical polar coordinates $(t, r, \vartheta, \varphi)$. Here, as in Eq.~\eqref{M15}, we have $P = a(t) - \phi$, $Q = a(t) + \phi$ and $\mathfrak{R} = (r^2+b^2)^{1/2}$, 
where $b$ is the Plummer radius and $\phi = M/(2\,\mathfrak{R})$.

The nonzero connection coefficients for metric~\eqref{A1} are given, up to the symmetry of the Christoffel symbols, by
\begin{equation}\label{A2}
\begin{split}
\Gamma^{t}_{tt} &= H\, \frac{2 a \phi}{PQ}\,, \quad 
\Gamma^{t}_{tr} = \frac{ M r a }{PQ\mathfrak{R}^{3}}\,,
\quad
r^2\,\Gamma^{t}_{rr} = \Gamma^{t}_{\vartheta \vartheta} = H\,\frac{r^2 Q^5}{a^2 P}\,, \quad
\Gamma^{t}_{\varphi \varphi} = \sin^2{\vartheta}\,\Gamma^{t}_{\vartheta \vartheta}\,,
\\
\Gamma^{r}_{\varphi \varphi} &= \sin^2{\vartheta}\,\Gamma^{r}_{\vartheta \vartheta}\,, \quad \Gamma^{r}_{tt} = \frac{Mra^3P}{Q^7\mathfrak{R}^{3}}\,, \quad 
\Gamma^{r}_{tr} = \Gamma^{\vartheta}_{t\vartheta} = \Gamma^{\varphi}_{t\varphi} = H\, \frac{P}{Q} \,, \quad \Gamma^{r}_{rr}  =  - \frac{Mr}{Q\mathfrak{R}^{3}}\,, 
\\
\Gamma^{r}_{\vartheta \vartheta}  & = -r + \frac{r^3 M}{Q\mathfrak{R}^{3}}\,, \quad \Gamma^{\vartheta}_{r\vartheta} = \Gamma^{\varphi}_{r\varphi} = -\frac{1}{r^2}\,\Gamma^{r}_{\vartheta \vartheta}\,, \quad \Gamma^{\vartheta}_{\varphi \varphi} = - \cos{\vartheta} \sin{\vartheta}\,, \quad  \Gamma^{\varphi}_{\vartheta \varphi} = \cot{\vartheta} \,.
\end{split}
\end{equation} 

\section{Comparison with the LTB model}

The McVittie and McVittie-Plummer spacetimes do not allow the radial infall of cosmic matter onto their respective inhomogeneities. To maintain this particular feature of these cosmological models, their perfect fluid sources have pressure and the corresponding pressure gradients are indeed required to prevent the radial infall of the fluid onto the inhomogeneities. Moreover, spacetime singularities occur where the pressure diverges. The LTB model, by contrast, is a spherically symmetric dust model discovered by Lema\^itre~\cite{Lemaitre:1933gd} and later studied by Tolman~\cite{Tolman:1934za} and Bondi~\cite{Bondi:1947fta}. In the LTB model, a spacetime singularity (including shell-crossing singularity) occurs where the energy density diverges~\cite{Krasinski:1997yxj, Plebanski:2006sd, Bolejko:2009pvd}.

The LTB model is based on a metric of the form~\eqref{R9} in comoving coordinates with 
\begin{equation}\label{B1}
A(t, r) = 1, \qquad B(t, r) = [1+ 2\,\mathcal{E}(r)]^{-1/2} \,\frac{\partial R}{\partial r}\, 
\end{equation}
and an energy-momentum tensor 
\begin{equation}\label{B2}
T_{\mu \nu} = \mu_{\rm LTB}(t, r)\, u_\mu u_\nu\,, \qquad u^\mu = \delta^\mu_0\,.
\end{equation}
The gravitational field equations~\eqref{R3} in this case imply
\begin{equation}\label{B3}
\mathcal{E}(r) = \frac{1}{2} \left(\frac{\partial R}{\partial t}\right)^2 -\frac{G\mathcal{M}(r)}{R} -\frac{1}{6} \Lambda R^2\,,
\end{equation}
\begin{equation}\label{B4}
\frac{d\mathcal{M}(r)}{dr} = 4 \pi \mu_{\rm LTB}(t, r)\,R^2 \,\frac{\partial R}{\partial r}\,,
\end{equation}
where $\mathcal{E}(r)$, $\mathcal{E}(r) > -\tfrac{1}{2}$,  is the energy per unit mass of a spherical shell of dust of radius $r$ and $\mathcal{M}(r)$ is the mass-energy within a sphere of radius $r$. It is straightforward to employ Eq.~\eqref{R11} to show that the Misner-Sharp mass in this case is independent of time and is indeed given by
\begin{equation}\label{B5}
m_{\rm LTB} = \mathcal{M}(r).
\end{equation}
The preferred comoving observers in the LTB model follow timelike geodesics of the spacetime manifold. 

The LTB model is a generalization of the standard spatially homogeneous and isotropic Friedmann model (i.e., FLRW universe with zero pressure). Let us first note that the metric of the FLRW universe can be expressed as 
\begin{equation}\label{B6}
ds^2|_{\rm FLRW} = -dt^2 + a^2(t) \left[\frac{dr^2}{1-\tfrac{k}{R_0^2}\,r^2} + r^2 d\Omega^2\right].
\end{equation}
The LTB metric reduces to the above metric for 
\begin{equation}\label{B7}
R(t, r) = a(t)\, r\,, \qquad 2\,\mathcal{E} = -\frac{k}{R_0^2}\,r^2\,, \qquad \mathcal{M} = \mathcal{M}_0\, r^3\,.
\end{equation}
In this case, $\mu_{\rm LTB}(t, r) \to \mu_{\rm F}(t)$, where $\mu_{\rm F}(t)$ is the energy density of the Friedmann universe, and $\mathcal{M}_0$ is a constant given by
\begin{equation}\label{B8}
 \mathcal{M}_0 = \frac{4 \pi}{3}\, \mu_{\rm F} \,a^3(t)\,.
\end{equation}
The vanishing of pressure in the Friedmann model implies that the energy density $\mu_{\rm F}$ is a constant divided by the scale parameter to the third power. 

Let us briefly digress here and mention that with 
\begin{equation}\label{B9}
r = \frac{\varpi}{1+\tfrac{k}{4R_0^2}\,\varpi^2}\,,
\end{equation}
metric~\eqref{B6} becomes
\begin{equation}\label{B10}
ds^2|_{\rm FLRW} = -dt^2 + \frac{a^2(t)}{(1+\tfrac{k}{4R_0^2}\,\varpi^2)^2} (d\varpi^2 + \varpi^2 d\Omega^2)\,,
\end{equation}
which is essentially equivalent, via a simple coordinate transformation, to the isotropic form of the FLRW metric given in Eq.~\eqref{H1}. 

Returning to the LTB model, we note that to have a proper physical interpretation, we must assume that as we move away from the central inhomogeneity and as $r$ increases beyond a certain radial coordinate, radial inhomogeneity gradually decreases and the LTB model approaches a Friedmann model. Let us clarify this situation with an example. A simple LTB model is given by 
\begin{equation}\label{B11}
R(t, r) = \left(\tfrac{9G}{2}\right)^{1/3}\mathcal{M}^{1/3}(r) [t - t_B(r)]^{2/3}\,,
\end{equation}
which is the general solution of Eq.~\eqref{B3} for $\mathcal{E} = \Lambda = 0$. Here, $t_B(r)$ specifies the time of the occurrence of the big bang singularity for a given radial coordinate $r$. The energy density $\mu_{\rm LTB}(t, r)$ can be calculated using Eq.~\eqref{B4} in terms of $\mathcal{M}(r)$, $t_B(r)$, and their spatial derivatives. Let us assume that as $r \to \infty$, $\mathcal{M} \to \mathcal{M}_0\, r^3$, where $\mathcal{M}_0$ is a constant, and $t_B(r)$ rapidly decreases with increasing $r$ such that
\begin{equation}\label{B12}
\lim_{r \to \infty} r \,\frac{dt_B(r)}{dr} \to 0\,; 
\end{equation}
then, we find that the spherical inhomogeneity is embedded in a flat Friedmann model known as the Einstein-de Sitter universe. 

The LTB model thus represents a spherically symmetric inhomogeneity embedded in a Friedmann background, just as the McVittie-Plummer model represents a spherical star system embedded in a FLRW background. 

In 1975,  Szekeres discovered a large class of irrotational dust cosmological models~\cite{Szekeres:1974ct}. The Szekeres metric~\cite{Krasinski:1997yxj, Plebanski:2006sd, Bolejko:2009pvd}, which is a generalization of LTB metric, has no symmetry in its general form and its consideration is beyond the scope of the present paper.

%%%%%%%%%%%
%%%%%%%%%%%
%%%%%%%%%%%


\begin{thebibliography}{00}

%\cite{Cooperstock:1998ny}
\bibitem{Cooperstock:1998ny}
F.~I.~Cooperstock, V.~Faraoni and D.~N.~Vollick,
``The Influence of the cosmological expansion on local systems",
Astrophys. J. \textbf{503}, 61 (1998).
%doi:10.1086/305956
[arXiv:astro-ph/9803097 [astro-ph]]
%89 citations counted in INSPIRE as of 31 May 2024


%\cite{Mashhoon:2007qm}
\bibitem{Mashhoon:2007qm}
B.~Mashhoon, N.~Mobed and D.~Singh,
``Tidal dynamics in cosmological spacetimes",
Classical Quantum Gravity \textbf{24}, 5031-5046 (2007).
%doi:10.1088/0264-9381/24/20/008
[arXiv:0705.1312 [gr-qc]]
%36 citations counted in INSPIRE as of 25 May 2024


%\cite{Faraoni:2007es}
\bibitem{Faraoni:2007es}
V.~Faraoni and A.~Jacques,
``Cosmological expansion and local physics",
Phys. Rev. D \textbf{76}, 063510 (2007).
%doi:10.1103/PhysRevD.76.063510
[arXiv:0707.1350 [gr-qc]]
%152 citations counted in INSPIRE as of 31 May 2024


%\cite{Kopeikin:2012by}
\bibitem{Kopeikin:2012by}
S.~Kopeikin,
``Celestial Ephemerides in an Expanding Universe",
Phys. Rev. D \textbf{86}, 064004 (2012).
%doi:10.1103/PhysRevD.86.064004
[arXiv:1207.3873 [gr-qc]]
%33 citations counted in INSPIRE as of 31 May 2024


%\cite{Kopeikin:2013am}
\bibitem{Kopeikin:2013am}
S.~M.~Kopeikin and A.~N.~Petrov,
``Post-Newtonian Celestial Dynamics in Cosmology: Field Equations",
Phys. Rev. D \textbf{87}, no.4, 044029 (2013).
%doi:10.1103/PhysRevD.87.044029
[arXiv:1301.5706 [gr-qc]]
%16 citations counted in INSPIRE as of 31 May 2024


%\cite{Iorio:2012wva}
\bibitem{Iorio:2012wva}
L.~Iorio,
``Local cosmological effects of order H in the orbital motion of a binary system?",
Mon. Not. Roy. Astron. Soc. \textbf{429}, 915-922 (2013).
%doi:10.1093/mnras/sts396
[arXiv:1208.1523 [gr-qc]]
%13 citations counted in INSPIRE as of 31 May 2024

%\cite{Spengler:2021vxy}
\bibitem{Spengler:2021vxy}
F.~Spengler, A.~Belenchia, D.~R\"atzel and D.~Braun,
``Influence of cosmological expansion in local experiments",
Classical Quantum Gravity \textbf{39}, no.5, 055005 (2022).
%doi:10.1088/1361-6382/ac4954
[arXiv:2109.03280 [gr-qc]]
%1 citations counted in INSPIRE as of 31 May 2024



%\cite{Nandra:2011ug}
\bibitem{Nandra:2011ug}
R.~Nandra, A.~N.~Lasenby and M.~P.~Hobson,
``The effect of a massive object on an expanding universe",
Mon. Not. Roy. Astron. Soc. \textbf{422}, 2931-2944 (2012).
%doi:10.1111/j.1365-2966.2012.20618.x
[arXiv:1104.4447 [gr-qc]]
%72 citations counted in INSPIRE as of 30 May 2024

%\cite{Nandra:2011ui}
\bibitem{Nandra:2011ui}
R.~Nandra, A.~N.~Lasenby and M.~P.~Hobson,
``The effect of an expanding universe on massive objects",
Mon. Not. Roy. Astron. Soc. \textbf{422}, 2945-2959 (2012).
%doi:10.1111/j.1365-2966.2012.20617.x
[arXiv:1104.4458 [gr-qc]]
%61 citations counted in INSPIRE as of 30 May 2024

%\cite{Nandra:2013jga}
\bibitem{Nandra:2013jga}
R.~Nandra, A.~Lasenby and M.~Hobson,
``Dynamics of a spherical object of uniform density in an expanding universe",
Phys. Rev. D \textbf{88}, no.4, 044041 (2013).
%doi:10.1103/PhysRevD.88.044041
[arXiv:1307.0526 [astro-ph.CO]]
%3 citations counted in INSPIRE as of 30 May 2024

%\cite{Benisty:2024tlv}
\bibitem{Benisty:2024tlv}
D.~Benisty, M.~M.~Chaichian and A.~Tureanu,
``Galaxy groups in the presence of cosmological constant: Increasing the masses of groups",
Phys. Lett. B \textbf{858}, 139033 (2024).
%doi:10.1016/j.physletb.2024.139033
[arXiv:2405.14944 [astro-ph.GA]]
%2 citations counted in INSPIRE as of 24 Oct 2024


\bibitem{Plummer}
H. C. Plummer, 
``On the problem of distribution in globular star clusters",
Mon. Not. R. Astron. Soc. \textbf{71}, 460-470 (1911).

%\cite{McVittie:1933zz}
\bibitem{McVittie:1933zz}
G.~C.~McVittie,
``The mass-particle in an expanding universe",
Mon. Not. Roy. Astron. Soc. \textbf{93}, 325-339 (1933).
%\url{https://doi:10.1093/mnras/93.5.325}
%374 citations counted in INSPIRE as of 12 Oct 2023

%\cite{Krasinski:1997yxj}
\bibitem{Krasinski:1997yxj}
A.~Krasinski,
\emph{Inhomogeneous Cosmological Models}
(Cambridge University Press, 2006).
%ISBN 978-0-511-88754-3, 978-0-521-48180-9, 978-0-521-03017-5
%37 citations counted in INSPIRE as of 30 May 2024



%\cite{Plebanski:2006sd}
\bibitem{Plebanski:2006sd}
J.~Plebanski and A.~Krasinski,
\emph{An introduction to general relativity and cosmology}, 2nd edn
(Cambridge University Press, 2024).
%41 citations counted in INSPIRE as of 30 May 2024


%\cite{Bolejko:2009pvd}
\bibitem{Bolejko:2009pvd}
K.~Bolejko, A.~Krasinski, C.~Hellaby and M.~N.~Celerier,
\emph{Structures in the Universe by Exact Methods}
(Cambridge University Press, 2009).
%doi:10.1017/CBO9780511657405
%8 citations counted in INSPIRE as of 30 May 2024

%\cite{Misner:1964je}
\bibitem{Misner:1964je}
C.~W.~Misner and D.~H.~Sharp,
``Relativistic equations for adiabatic, spherically symmetric gravitational collapse",
Phys. Rev. \textbf{136}, B571-B576 (1964). 
%\url{https://doi:10.1103/PhysRev.136.B571}
%1047 citations counted in INSPIRE as of 02 Feb 2024

\bibitem{HeMi}
W. C. Hernandez, Jr. and C. W. Misner,
``Observer time as a coordinate in relativistic spherical hydrodynamics",
Astrophys. J. \textbf{143}, 452-464 (1966).

\bibitem{CaMc}
M. E. Cahill and G. C. McVittie,
``Spherical Symmetry and Mass-Energy in General Relativity. I. General Theory", 
J. Math. Phys. \textbf{11}, 1382-1391 (1970).
%https://doi.org/10.1063/1.1665273


\bibitem{Bonnor}
W. B. Bonnor, 
`` Local Dynamics and the Expansion of the Universe", 
Gen. Relativ. Gravit. \textbf{32}, 1005-1007 (2000).
%DOI: 10.1023/A:1001961325184 


%\cite{Kaloper:2010ec}
\bibitem{Kaloper:2010ec}
N.~Kaloper, M.~Kleban and D.~Martin,
``McVittie's Legacy: Black Holes in an Expanding Universe",
Phys. Rev. D \textbf{81}, 104044 (2010).
%\url{https://doi:10.1103/PhysRevD.81.104044},
[arXiv:1003.4777 [hep-th]]
%136 citations counted in INSPIRE as of 12 Oct 2023

%\cite{Lake:2011ni}
\bibitem{Lake:2011ni}
K.~Lake and M.~Abdelqader,
``More on McVittie's Legacy: A Schwarzschild - de Sitter black and white hole embedded in an asymptotically $\Lambda$CDM cosmology",
Phys. Rev. D \textbf{84}, 044045 (2011).
%\url{https://doi:10.1103/PhysRevD.84.044045},
[arXiv:1106.3666 [gr-qc]]
%78 citations counted in INSPIRE as of 12 Oct 2023

%\cite{Nolan:2014maa}
\bibitem{Nolan:2014maa}
B.~C.~Nolan,
``Particle and photon orbits in McVittie spacetimes",
Classical Quantum Gravity \textbf{31}, no.23, 235008 (2014).
%doi:10.1088/0264-9381/31/23/235008
[arXiv:1408.0044 [gr-qc]]
%34 citations counted in INSPIRE as of 30 May 2024

%\cite{Nolan:2017rtj}
\bibitem{Nolan:2017rtj}
B.~C.~Nolan,
``Local properties and global structure of McVittie spacetimes with non-flat Friedmann\textendash{}Lema\^\i{}tre\textendash{}Robertson\textendash{}Walker backgrounds",
Classical Quantum Gravity \textbf{34}, no.22, 225002 (2017).
%\url{https://doi:10.1088/1361-6382/aa903c},
[arXiv:1707.07612 [gr-qc]]
%8 citations counted in INSPIRE as of 12 Oct 2023

%\cite{Perlick:2018iye}
\bibitem{Perlick:2018iye}
V.~Perlick, O.~Y.~Tsupko and G.~S.~Bisnovatyi-Kogan,
``Black hole shadow in an expanding universe with a cosmological constant",
Phys. Rev. D \textbf{97}, no.10, 104062 (2018).
%\url{https://doi:10.1103/PhysRevD.97.104062},
[arXiv:1804.04898 [gr-qc]]
%134 citations counted in INSPIRE as of 12 Oct 2023

%\cite{Faraoni:2018xwo}
\bibitem{Faraoni:2018xwo}
V.~Faraoni,
``Embedding black holes and other inhomogeneities in the universe in various theories of gravity: a short review",
Universe \textbf{4}, no.10, 109 (2018).
%\url{https://doi:10.3390/universe4100109},
[arXiv:1810.04667 [gr-qc]]
%27 citations counted in INSPIRE as of 12 Oct 2023

%\cite{Gaur:2023hmk}
\bibitem{Gaur:2023hmk}
R.~Gaur and M.~Visser,
``Black holes embedded in FLRW cosmologies",
Phys. Rev. D \textbf{110}, no.4, 043529 (2024).
%doi:10.1103/PhysRevD.110.043529
[arXiv:2308.07374 [gr-qc]]
%13 citations counted in INSPIRE as of 24 Oct 2024

\bibitem{BuD}
H. A. Buchdahl, 
``A relativistic fluid sphere resembling the Emden polytrope of index 5", 
Astrophys. J. \textbf{140}, 1512-1516 (1964).

\bibitem{Facke}
E. D. Fackerell, 
``Relativistic, spherically symmetric star clusters. V. A relativistic version of Plummer's model",
Astrophys. J. \textbf{165}, 489-493 (1971).

\bibitem{GlaMa}
E. N. Glass and B. Mashhoon, 
``On a spherical star system with a collapsed core", 
Astrophys. J. \textbf{205}, 570-577 (1976).

\bibitem{Einstein} 
A. Einstein, \emph{The Meaning of Relativity}
(Princeton University Press, Princeton, NJ, USA, 1955). 




%\cite{Mashhoon:1979tt}
\bibitem{Mashhoon:1979tt}
B.~Mashhoon and M.~H.~Partovi,
``Gravitational Collapse of a Charged Fluid Sphere",
Phys. Rev. D \textbf{20}, 2455-2468 (1979).
%\url{https://doi:10.1103/PhysRevD.20.2455}
%30 citations counted in INSPIRE as of 09 Oct 2023

%\cite{Dodelson:2003ft}
\bibitem{Dodelson:2003ft}
S.~Dodelson,
{\it{Modern Cosmology}} (Academic Press, 2003).
%ISBN 978-0-12-219141-1
%419 citations counted in INSPIRE as of 04 Jun 2024


%\cite{Weinberg:2008zzc}
\bibitem{Weinberg:2008zzc}
S.~Weinberg, {\it{Cosmology}} (Oxford University Press, 2008).
%525 citations counted in INSPIRE as of 04 Jun 2024

%\cite{Amendola:2015ksp}
\bibitem{Amendola:2015ksp}
L.~Amendola and S.~Tsujikawa,
{\it{Dark Energy: Theory and Observations}} (Cambridge University Press, 2015).
%ISBN 978-1-107-45398-2
%14 citations counted in INSPIRE as of 04 Jun 2024

%\cite{Baumann:2022mni}
\bibitem{Baumann:2022mni}
D.~Baumann, {\it{Cosmology}} (Cambridge University Press, 2022).
%ISBN 978-1-108-93709-2, 978-1-108-83807-8
%doi:10.1017/9781108937092
%46 citations counted in INSPIRE as of 04 Jun 2024

\bibitem{HE}
S. W. Hawking and G. F. R. Ellis, 
\emph{The Large Scale Structure of Space-Time}
(Cambridge University Press, Cambridge, UK, 1973).

\bibitem{Kottler}
F. Kottler, 
``\"Uber die physikalischen Grundlagen der Einsteinschen Gravitationstheorie",
Ann. Phys. (Leipzig)  \textbf{56}, 410-462 (1918). 
% https://doi.org/10.1002/andp.19183611402


%\cite{Chicone:2005vn}
\bibitem{Chicone:2005vn}
C.~Chicone and B.~Mashhoon,
``Explicit Fermi coordinates and tidal dynamics in de Sitter and G\"odel spacetimes",
Phys. Rev. D \textbf{74}, 064019 (2006).
%doi:10.1103/PhysRevD.74.064019
[arXiv:gr-qc/0511129 [gr-qc]]
%61 citations counted in INSPIRE as of 24 May 2024

\bibitem{Mashhoon}
B. Mashhoon, 
``Is the Universe Homogeneous on a Large Scale?", 
in: \emph{The Big Bang and Georges Lema\^itre}, edited by A. Berger (Reidel, Dordrecht, 1984), pp. 75-81. 



%\cite{Cadoni:2023lqe}
\bibitem{Cadoni:2023lqe}
M.~Cadoni, R.~Murgia, M.~Pitzalis and A.~P.~Sanna,
``Quasi-local masses and cosmological coupling of black holes and mimickers",
JCAP \textbf{03}, 026 (2024).
%doi:10.1088/1475-7516/2024/03/026
[arXiv:2309.16444 [gr-qc]]
%3 citations counted in INSPIRE as of 04 Apr 2024


%\cite{Cadoni:2024jxy}
\bibitem{Cadoni:2024jxy}
M.~Cadoni, M.~Pitzalis, D.~C.~Rogridues and A.~P.~Sanna,
``Cosmological coupling of local gravitational systems",
[arXiv:2406.06091 [gr-qc]].
%0 citations counted in INSPIRE as of 12 Jun 2024

%\cite{Gunn:1972sv}
\bibitem{Gunn:1972sv}
J.~E.~Gunn and J.~R.~Gott, III,
``On the Infall of Matter into Clusters of Galaxies and Some Effects on Their Evolution",
Astrophys. J. \textbf{176}, 1-19 (1972).
%doi:10.1086/151605
%2283 citations counted in INSPIRE as of 15 Aug 2024

%\cite{Schneider:2006qyj}
\bibitem{Schneider:2006qyj}
P.~Schneider, C.~S.~Kochanek and J.~Wambsganss,
``Proceedings, 33rd Advanced Saas Fee Course on Gravitational Lensing: Strong, Weak, and Micro: Les Diablerets, Switzerland, April 7-12, 2003",
Saas-Fee Advanced Courses \textbf{33}, pp.1-553 (2006).
%14 citations counted in INSPIRE as of 18 Aug 2024


%\cite{Saha:2024axf}
\bibitem{Saha:2024axf}
P.~Saha, D.~Sluse, J.~Wagner and L.~L.~R.~Williams,
``Essentials of Strong Gravitational Lensing",
Space Sci. Rev. \textbf{220}, no.1, 12 (2024).
%doi:10.1007/s11214-024-01041-w
[arXiv:2401.04165 [astro-ph.CO]]
%6 citations counted in INSPIRE as of 01 Aug 2024


%\cite{Piattella:2015xga}
\bibitem{Piattella:2015xga}
O.~F.~Piattella,
``Lensing in the McVittie metric",
Phys. Rev. D \textbf{93}, no.2, 024020 (2016),
[erratum: Phys. Rev. D \textbf{93}, no.12, 129901 (2016)].
%doi:10.1103/PhysRevD.93.024020
[arXiv:1508.04763 [astro-ph.CO]]
%26 citations counted in INSPIRE as of 01 Aug 2024

%\cite{Piattella:2016nzt}
\bibitem{Piattella:2016nzt}
O.~F.~Piattella,
``On the effect of the cosmological expansion on the gravitational lensing by a point mass",
Universe \textbf{2}, no.4, 25 (2016).
%doi:10.3390/universe2040025
[arXiv:1609.00270 [gr-qc]]
%12 citations counted in INSPIRE as of 18 Aug 2024

 
%\cite{Bessa:2022sdh}
\bibitem{Bessa:2022sdh}
P.~Bessa and O.~F.~Piattella,
``Gravitational lensing in a universe with matter and a cosmological constant",
Phys. Rev. D \textbf{106}, no.12, 123513 (2022).
%doi:10.1103/PhysRevD.106.123513
[arXiv:2209.04063 [astro-ph.CO]]
%3 citations counted in INSPIRE as of 16 Aug 2024

 
 
%\cite{Persic:1995ru}
\bibitem{Persic:1995ru}
M.~Persic, P.~Salucci and F.~Stel,
``The universal rotation curve of spiral galaxies: 1. The dark matter connection",
Mon. Not. Roy. Astron. Soc. \textbf{281}, 27 (1996).
%doi:10.1093/mnras/278.1.27
[arXiv:astro-ph/9506004 [astro-ph]]
%972 citations counted in INSPIRE as of 01 Aug 2024

%\cite{Dubinski:1991bm}
\bibitem{Dubinski:1991bm}
J.~Dubinski and R.~G.~Carlberg,
``The Structure of cold dark matter halos",
Astrophys. J. \textbf{378}, 496 (1991).
%doi:10.1086/170451
%760 citations counted in INSPIRE as of 01 Aug 2024

%\cite{Noyola:2006mg}
\bibitem{Noyola:2006mg}
E.~Noyola and K.~Gebhardt,
``Surface Brightness Profiles of Galactic Globular Clusters from Hubble Space Telescope Images",
Astron. J. \textbf{132}, 447-466 (2006).
%doi:10.1086/505390
[arXiv:astro-ph/0604251 [astro-ph]]
%65 citations counted in INSPIRE as of 15 Aug 2024


%\cite{Lemaitre:1933gd}
\bibitem{Lemaitre:1933gd}
G.~Lema\^itre,
``L'univers en expansion",
Annales Soc. Sci. Bruxelles A \textbf{53}, 51-85 (1933).
%doi:10.1023/A:1018855621348
[Reprinted with English Translation: ``The expanding universe", Gen. Relativ. Gravit. \textbf{29}, 641-680 (1997).]
%764 citations counted in INSPIRE as of 18 Jul 2024

%\cite{Tolman:1934za}
\bibitem{Tolman:1934za}
R.~C.~Tolman,
``Effect of inhomogeneity on cosmological models",
Proc. Nat. Acad. Sci. \textbf{20}, 169-176 (1934).
%doi:10.1073/pnas.20.3.169
[Reprinted: Gen. Relativ. Gravit. \textbf{29}, 935 (1997).]
%857 citations counted in INSPIRE as of 18 Jul 2024

%\cite{Bondi:1947fta}
\bibitem{Bondi:1947fta}
H.~Bondi,
``Spherically symmetrical models in general relativity",
Mon. Not. Roy. Astron. Soc. \textbf{107}, 410-425 (1947).
%doi:10.1093/mnras/107.5-6.410
[Reprinted: Gen. Relativ. Gravit. \textbf{31}, 1777 (1999).]
%744 citations counted in INSPIRE as of 18 Jul 2024

%\cite{Szekeres:1974ct}
\bibitem{Szekeres:1974ct}
P.~Szekeres,
``A Class of Inhomogeneous Cosmological Models",
Commun. Math. Phys. \textbf{41}, 55-64 (1975).
%doi:10.1007/BF01608547
%324 citations counted in INSPIRE as of 19 Jul 2024


\end{thebibliography}
\end{document}